# Brillouin microscopy, what is it really measuring?

Pei-Jung Wu[1,2], Irina Kabakova[2#], Jeffrey Ruberti[3], Joseph M. Sherwood[1], Iain E. Dunlop[4], Carl Paterson[2], Peter Török[2]*, Darryl R. Overby[1]*

[1] Imperial College London, Department of Bioengineering, London UK
[2] Imperial College London, Department of Physics, London UK
[3] Northeastern University, Department of Bioengineering, Boston, MA USA
[4] Imperial College London, Department of Materials, London UK

*co-corresponding authors

[#]current address: University Technology Sydney, School of Mathematical and Physical Sciences, Sydney, NSW, Australia

**Abstract**

Brillouin microscopy measures *compressibility*, but is being increasingly used to assess *stiffness* of cells and tissues. Using hydrogels with tunable properties, we demonstrate that Brillouin microscopy is insensitive to stiffness of hydrated materials, but depends strongly on water content, consistent with a theoretical model of biphasic compressibility. Empirical correlations between Brillouin measurements and stiffness arise due to their mutual dependence on water content, but correlations vanish once hydration is controlled.

**Main Text**

Brillouin microscopy is an optical method to putatively map the mechanical properties of materials using a confocal microscope[1-3]. The first Brillouin images of cells were reported in 2015[4,5] and inspired a growing number of applications in cardiovascular disease[6], ophthalmology[7,8], and cellular mechanics[4,5,9-12]. Conventional micromechanical tools, such as atomic force microscopy, are often destructive, require contact and provide only surface measurements. Brillouin microscopy, in contrast, offers non-invasive, label-free access to the internal micromechanics of cells and tissues at a resolution unachievable by current forms of ultrasound or optical elastography[13].

Brillouin microscopy relies on the phenomenon of Brillouin scattering[14], whereby photons exchange energy with thermally-driven acoustic waves, or phonons, leading to a frequency shift $\omega_b$ between the incident and scattered light, given by

$$\omega_b = \frac{2n}{\lambda}\sqrt{\frac{M}{\rho}}\,\sin\frac{\theta}{2} \qquad \text{Eq. 1}$$

$\rho$ and $n$ are the density and refractive index of the material, and $\lambda$ and $\theta$ are the *in vacuo* wavelength and angle between incident and scattered wave vectors. $M$ is the longitudinal elastic modulus that represents the compressibility of a material, or specifically the mechanical stress necessary to impose a volume change in the material by compressing or expanding it in one direction. $\lambda$ and $\theta$ are typically prescribed by the setup of the Brillouin microscope, which measures $\omega_b$, yielding an estimate of $M$ once $\rho$ and $n$ are determined.

Despite Equation 1 depending on $M$, recent reports have interpreted Brillouin micrographs in terms of the Young's modulus $E$ for cells and tissues[3,4,10]. $E$ represents the stiffness of a material, i.e. how resistive it is to deformation, and describes the stress necessary to compress or extend a material in one direction without constraining the volume. As biological materials are composed mostly of water, which is relatively incompressible, $M$ is several orders of magnitude greater than $E$ ($10^9$ vs. $10^2$–$10^6$ Pa). Further, Brillouin scattering is sensitive to GHz frequencies where, due to viscoelastic effects, measurements may diverge from those measured at lower strain rates. As physiological deformation typically occurs relatively slowly with negligible changes in volume, cell and tissue mechanics depend on $E$ and are largely insensitive to $M$. Nonetheless, empirical correlations between $M$ and $E$ for cells[4], hydrogels[3,4] and other biological tissues[3] have suggested that variations in $M$, as measured by Brillouin microscopy, reflect variations in $E$. Based on these correlations, authors have proposed equations directly relating $M$ and $E$, implying that Brillouin measurements reflect cell or tissue stiffness[3,4,10]. Being empirical, these correlations do not necessarily imply a direct relationship between $M$ and $E$. As water content can affect both $M$ and $E$ in hydrated materials, we hypothesize that Brillouin microscopy does not measure $E$ directly, but more likely reflects the volume fraction $\varepsilon$ of water contained within the hydrated material.

Motivated by the potential for Brillouin microscopy as a revolutionary tool for cellular mechanobiology, we set out to examine more closely the relationship between Brillouin microscopy and Young's modulus in hydrated materials, accounting for the potential influence of water content. We used hydrogels as a simple model for biological materials because both contain fluid interspersed within a flexible solid network that provides elasticity.

With increasing molecular weight of polyethylene oxide (PEO), PEO hydrogels change from a dilute suspension with zero Young's modulus to a semi-dilute entangled network with finite $E$[15]. Further, by decreasing the polymer concentration in proportion to the increase in molecular weight, water content $\varepsilon$ can be fixed whilst increasing $E$. Thus, we could vary $E$ and $\varepsilon$ independently whilst measuring $M$. With increasing molecular weight, $E$ measured by rheometry increased for a given $\varepsilon$ (Figure 1A). However, $M$ was unaffected by the change in molecular weight, but decreased almost linearly with $\varepsilon$ (Figure 1B). Thus, for PEO hydrogels, changes in $E$ were uncorrelated with changes in $M$ when controlling for water content (Figure 1C). Similar results were observed in a second experiment (Supplemental Note 1).

To understand the relationship between $M$ and $\varepsilon$, we consider acoustic wave propagation through a hydrogel where both the fluid and solid components transfer acoustic momentum. The aggregate compressibility of the biphasic medium is then equal to the sum of the individual fluid and solid compressibilities weighted by their respective volume fractions[16,17]:

$$\frac{1}{M} = \frac{\varepsilon}{M_f} + \frac{1-\varepsilon}{M_s} \qquad \text{Eq. 2}$$

where $M_f$ and $M_s$ are the longitudinal elastic moduli (inverse compressibilities) of the fluid and solid, respectively. Because salinity affects $M_f$, we then measured $M$ for PEO hydrogels over a range of NaCl concentrations and volume fractions of saline $\varepsilon$. The predictions of Equation 2 were consistent with the measured values of $M$ using a single value for $M_s$ and measured values of $M_f$ for each saline solution (Figure 1D).

Previous correlations between $M$ and $E$ were reported based on polyacrylamide (PA) hydrogels[4]. PA hydrogels swell, and thereby increase their water content over time (Supplemental Figure 1). We used several different concentrations of bis-acrylamide cross-linker to vary hydrogel stiffness, and measured the longitudinal elastic modulus, Young's modulus and water content during swelling to investigate the relationships between $M$, $E$ and $\varepsilon$.

At any given time point during swelling, there was a strong relationship between $E$ measured by unconfined compression and $\varepsilon$ when comparing between hydrogels of different cross-linker concentrations (Figure 2A). However, when following individual hydrogels over time during swelling, $E$ remained nearly constant or decreased only slightly (Figure 2B). The change in $E$ over time was consistent with theoretical predictions of hydrogel swelling[18]:

$$E = E_0 \, Q^{-\frac{1}{3}} \qquad \text{Eq. 3}$$

where $Q = (1 - \varepsilon_0)/(1 - \varepsilon)$ is the swelling ratio and $E_0$ is the Young's modulus immediately after gelation when $\varepsilon = \varepsilon_0$.

$M$ also decreased during swelling for PA hydrogels, as predicted by Equation 2 whenever $\varepsilon$ increases with $M_s > M_f$. Regardless of the extent of swelling, all values of $M$ collapsed onto a single relationship consistent with Equation 2 when plotted versus $\varepsilon$, despite differences in $E$ or cross-linker concentration (Figure 2C). Thus, like PEO hydrogels, Brillouin microscopy of PA hydrogels appears highly sensitive to water content and relatively insensitive to $E$.

When $M$ was plotted versus $E$ for PA hydrogels, the resulting point cloud suggested a correlation when examining individual time points across multiple cross-linker concentrations or when aggregating all data together (Supplemental Note 2). However, when tracking individual hydrogels over time during swelling, no single trajectory captured how $M$ changed in relation to $E$ (Figure 2D). In contrast, Equations 2 and 3 predicted the trajectory of $M$ and $E$ for individual hydrogels as $\varepsilon$ changed over time during swelling, using single values of $M_s$ and $M_f$ for all cross-linker concentrations. These predicted trajectories (indicated by solid curves in Figure 2D) explain how a correlation between $M$ and $E$ can emerge and change over time for PA hydrogels, without an explicit dependence of $M$ upon $E$. Thus, while $M$ and $E$ may be correlated for PA hydrogels, this correlation is an epiphenomenon resulting from their mutual dependence on water content that changes during swelling. After accounting for water content, there is no obvious residual correlation between $M$ and $E$ (Supplemental Note 2). Prior correlations between $M$ and $E$ using PA[4,9] or biological materials[3,4] may have been similarly affected by water content.

In conclusion, Brillouin microscopy applied to soft hydrated materials measures a volume-weighted aggregate longitudinal elastic modulus where both fluid and solid components contribute to $M$. As biological materials are composed largely of water, hydration will tend to dominate biological Brillouin measurements and minimize the elastic contribution of the solid component. After correcting for the effect of hydration, the longitudinal elastic modulus measured by Brillouin microscopy does not necessarily correlate with the Young's modulus for hydrated materials. This work cautions against the straightforward use of Brillouin microscopy, or Brillouin scattering in general, as a form of optical elastography when applied to biological materials, but suggests that Brillouin microscopy may be well suited for investigating mechanisms involving local hydration, such as cell volume regulation, intracellular phase changes and polymerization.

**Methods**

Methods are given in the supplemental information.

# Author Contribution Statement

IK, CP, PT, JR, JMS, IED and DRO planned the study. PJW and IK conducted experiments. All authors participated and contributed to data analysis. DRO and PJW wrote the manuscript. All authors contributed to editing and revising the manuscript.

**Figures**

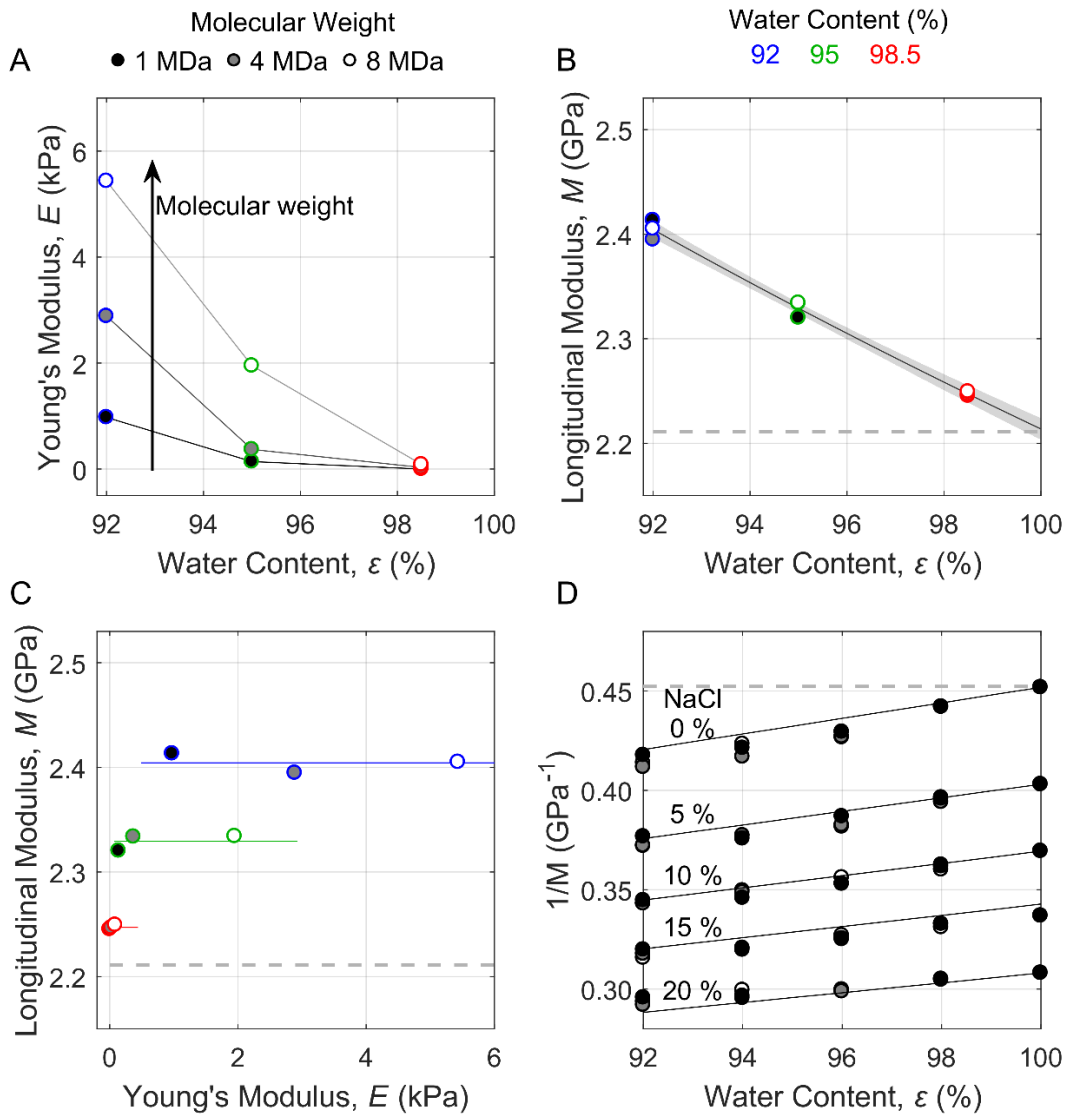

**Figure 1:** Polyethylene oxide hydrogels of varying molecular weight were used to examine independently the effect of Young's modulus $E$ and water content $\varepsilon$ on the longitudinal elastic modulus measured by Brillouin microscopy $M$. (A) With increasing molecular weight (arrow), $E$ measured by rheometry increased for a given value of $\varepsilon$ due to increasing entanglements between longer polymer chains. (B) Despite the changes in $E$, all values of $M$ collapsed onto a single curve predicted by Equation 2 when plotted versus $\varepsilon$ ($R^2$ = 0.97). Shaded regions represent 95% confidence bounds on the fit. (C) Increasing $E$ did not coincide with a change in $M$ after controlling for water content. Lines represent values of constant $\varepsilon$. (D) Increasing salinity from 0 to 20% NaCl in water increases $M_f$, yet the relationship between $M$ and $\varepsilon$ remains consistent with Equation 2. $M_f$ was measured for each saline concentration at $\varepsilon$ = 100%, and lines represent predictions from Equation 2 with $M_s$ = 16.35 GPa (see Supplemental Methods). Colors in panels A, B and C represent water content, with shading within the symbols representing molecular weight. Dashed horizontal lines in panels B, C and D represent the longitudinal elastic modulus of pure water. Data points correspond to individual hydrogels. Similar results were observed in a second experiment (Supplemental Note 1).

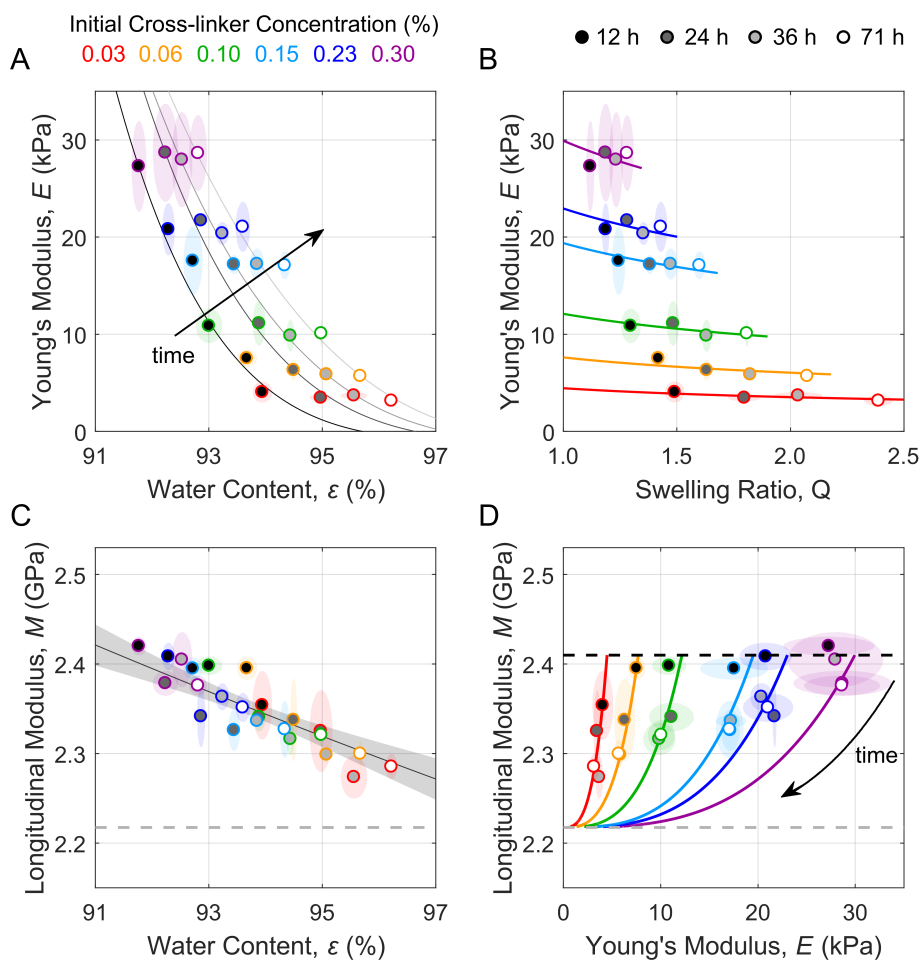

**Figure 2:** Polyacrylamide hydrogels of varying bis-acrylamide cross-linker concentration were allowed to swell in water, while measuring water content $\varepsilon$, Young's modulus $E$, and longitudinal elastic modulus by Brillouin microscopy $M$. (A) $E$ measured by unconfined compression decreased with increasing $\varepsilon$ when comparing between hydrogels of different cross-linker concentrations at any given time point during swelling (indicated by trend lines). (B) During swelling, $E$ for individual hydrogels remained constant or decreased only slightly, following the relationship $E \sim Q^{-1/3}$ as predicted by Equation 3, where $Q$ is the swelling ratio. (C) Brillouin data from all time points and cross-linker concentrations collapsed onto a single curve predicted by Equation 2 when plotted versus $\varepsilon$ ($R^2$ = 0.78). Shaded regions in panels C represent 95% confidence bounds. (D) No single relationship could describe how $M$ changed as a function of $E$ during swelling for individual hydrogels. However, changes in $M$ and $E$ during swelling were consistent with predictions of Equations 2 and 3 using $M_f$ = 2.21 GPa and $M_s$ = 16.35 GPa (curves). Arrow indicates the direction of increasing time during swelling. $E_0$ was estimated from Equation 3 using the fits shown in panel B. Colors represent different cross-linker concentrations. Grey dashed horizontal lines in panels C and D represent the longitudinal modulus of pure water. Black dashed line in panel D represents the estimated initial value of $M$ at the start of swelling when $\varepsilon_0$ was similar for all hydrogels. Elliptical regions surrounding the data points represent the two standard deviations on measurements from 4 replicate hydrogels. Similar results were observed in an additional experiment (Supplemental Note 1).

# Methods

## 1. Hydrogel Preparation

### 1.1 Polyethylene oxide (PEO)

PEO hydrogels were made from powder with an average molecular weight of 1, 4, and 8 MDa (Sigma-Aldrich). Mixtures of 1.5%, 5.0%, and 8.0% (v/v) PEO (corresponding to $\varepsilon$ = 0.985, 0.95 and 0.92, respectively) were prepared in distilled water with continuous magnetic stirring at 1000 rpm for at least 10 hours until the mixture was homogeneous. Data shown in the main text were from a single experiment with one hydrogel for each combination of water content and molecular weight. One sample volume from each hydrogel was taken for refractometry, one for Brillouin microscopy and 2-3 sample volumes for rheometry, as described below. PEO hydrogels were not allowed to swell and were mixed and stored in a sealed container to minimize evaporation. The density of PEO hydrogels was calculated according to:

$$\rho = \frac{m_w + m_p}{\frac{m_w}{\rho_w} + \frac{m_p}{\rho_p}} \quad \text{(S1)}$$

where $m_w$, $\rho_w$, $m_p$ and $\rho_p$ are the mass and density of water and polymer ($\rho_w$ = 1.00 g/cm$^3$, $\rho_p$ = 1.21 g/cm$^3$), respectively.

In a separate experiment, PEO hydrogels were prepared in a saline solvent to test the predictions of Equation 2 (Figure 1D). For this experiment, Mixtures of 2.0%, 4.0%, 6.0%, and 8.0% (v/v) PEO were prepared in 0%, 5%, 10%, 15%, and 20% (w/v) sodium chloride (NaCl) in distilled water. The preparation was otherwise identical to that described above. Hydrogel density was calculated as described in Equation S1. $\rho_w$ was measured by placing 8 known volumes (0.4 – 3.2 ml) on an analytical balance with a precision of 1 mg and performing a linear regression analysis between mass and volume, yielding $\rho_w$ = 1.00, 1.03, 1.06, 1.09 and 1.12 g/cm$^3$ for 0%, 5%, 10%, 15%, and 20% (w/v) NaCl, respectively. One sample volume from each hydrogel was taken to measure refractive index, one for Brillouin microscopy and 2-3 sample volumes for rheometry, as described below. The 0% NaCl condition represents a replicate of the experiment shown in Figures 1A-C of the main text where hydrogels were prepared in distilled water. These data are presented in Supplemental Note 1.

### 1.2 Polyacrylamide (PA)

PA hydrogels were prepared following a published protocol[1]. Hydrogel stiffness was adjusted by varying the ratio of acrylamide, bis-acrylamide and distilled water. For all samples, the initial concentration of acrylamide (Sigma-Aldrich) was 10% (w/v). N-methylene-bis-acrylamide (Sigma-Aldrich) was added at an initial concentration of 0.03, 0.06, 0.10, 0.15, 0.23, and 0.30% (w/v). The mixtures were centrifuged at 500g for 5 minutes, followed by the addition of 0.1% (w/v) ammonium persulfate (Sigma-Aldrich) and 0.1% tetramethylethylenediamine (Sigma-Aldrich) to initiate polymerization. The mixture was cast into disk-shaped molds (diameter 25 mm, depth 10 mm) and allowed to polymerize for 15 minutes. The hydrogels were then removed and immersed fully in a large container of distilled water to swell freely. Data shown in the main text were from a single experiment with 4 replicate hydrogels made from the same stock solution for each cross-linker concentration. A second experiment using the same preparation is shown in Supplemental Note 1.

Hydrogel mass $m$ was measured for each replicate immediately after polymerization and at 0, 12, 24, 36, and 71 hours using an electronic balance with a precision of 1 mg to track swelling (Figure

*Supplemental Information*

S1A). Brillouin microscopy, compression tests and refractometry were performed for each replicate at 12, 24, 36, and 71 hours after polymerization, as described below.

The water content $\varepsilon$ of PA hydrogels was calculated as

$$\varepsilon = \frac{V_w}{V} \qquad (S2)$$

where $V_w$ is the volume of water within the hydrogel and $V$ is the total hydrogel volume. $V_w$ is determined by the initial water volume at the start of polymerization, $V_{w,0}$, and the volume of water imbibed during swelling, $\Delta V_w$, such that $V_w = V_{w,0} + \Delta V_w$. Likewise, $V = V_0 + \Delta V_w$, where $V_0$ is the initial hydrogel volume determined by the volume of the mold. $V_{w,0}$ was calculated according to

$$V_{w,0} = V_0 - \frac{m_a}{\rho_a} - \frac{m_b}{\rho_b} \qquad (S3)$$

where $m_a$, $\rho_a$ and $m_b$, $\rho_b$ are the mass and density of acrylamide ($\rho_a$ = 1.13 g/cm³) and bis-acrylamide ($\rho_b$ = 1.235 g/cm³), respectively. $\Delta V_w$ was calculated according to

$$\Delta V_w = \frac{m - m_0}{\rho_w} \qquad (S4)$$

where $\rho_w$ is the density of water and $m_0$ is the hydrogel mass measured immediately after polymerization. The density of the hydrogel was calculated according to

$$\rho = \frac{m}{V} \qquad (S5)$$

The swelling ratio $Q$ of the hydrogel was calculated according to:

$$Q = \frac{V}{V_0} = \frac{1 - \varepsilon_0}{1 - \varepsilon} \qquad (S6)$$

where $\varepsilon_0 = V_{w,0}/V_0$ is the value of $\varepsilon$ immediately after polymerization (Figure S1B).

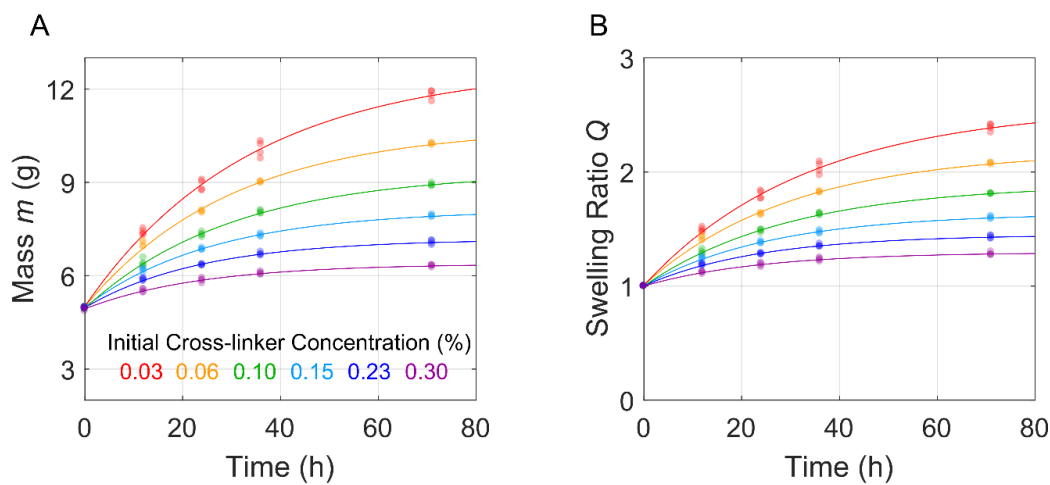

**Figure S1:** Polyacrylamide hydrogels of varying initial bis-acrylamide concentration during swelling. (A) Hydrogel mass $m$ increased during swelling, with larger swelling observed for lower cross-linker concentrations. (B) Swelling ratio $Q$ increased during swelling, as calculated by Eq. S6. Data are taken from the studies shown in Figure 2 of the main text. Each data point represents an individual hydrogel (*n* = 4) made from the same stock solution. Curves show exponential fits.

*Supplemental Information*

## 2. Brillouin microscopy

### 2.1 Experimental setup

Figure S2A shows the optical setup for Brillouin microscopy, as described in previous studies[2]. An objective lens (20X, NA = 0.5, Olympus) focused the light of a single longitudinal mode DPSS laser (λ=561nm; 30 mW, Cobolt Jive) emerging from single-mode optical fiber (Kineflex, OptiQ) to illuminate a volume of approximately 7 μm$^3$. Backscattered light from this volume was collected by the same objective and coupled into a single-mode optical fiber delivering the light to a custom-built spectrometer, consisting of an interferometer[2], a virtually imaged phased array (VIPA) and a Neo sCMOS camera (Andor). The interferometer suppresses Rayleigh peaks before the signal passes through the VIPA (LightMachinery Inc.) etalon that spatially separates the frequency components. The spectral resolution of the VIPA spectrometer used to produce the data in Figure 1A-C of the main text was 0.7 GHz with a finesse of 56 and an FSR of 39 GHz. For the data shown in Figure 1D and Figure 2 of the main text, the spectral resolution was 0.4 GHz with a finesse of 70 and an FSR of 30 GHz. For each PA or PEO hydrogel, three randomly-selected locations were measured in the hydrogel, with 50 spectra acquired at each location and averaged. The illumination and detection side single mode optical fibers together ensure that the optical system embodies a Type II confocal microscope[3]. This was important because measurements assume homogeneity within the sample volume.

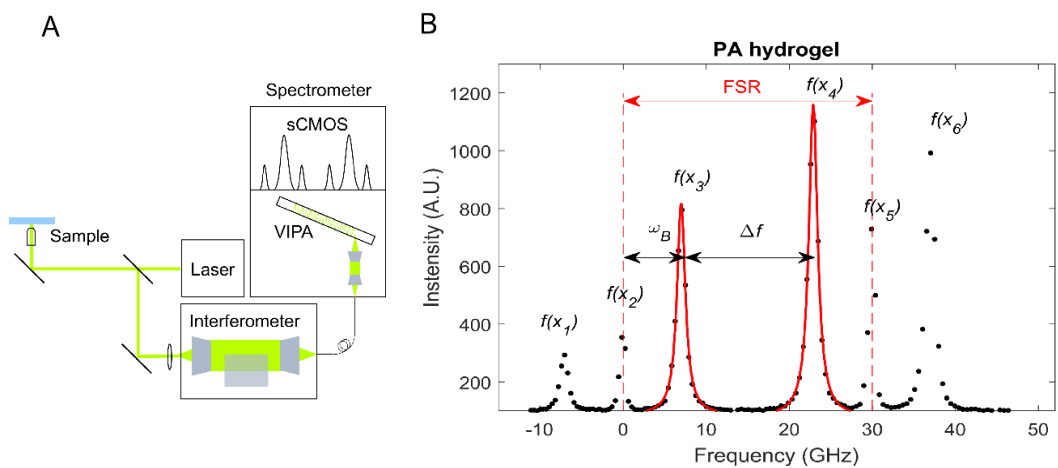

**Figure S2:** (A) Schematic of the Brillouin microscope. Laser light is directed into an inverted confocal microscope. Backscattered light is collected and filtered to reduce the intensity of the Rayleigh peak by up to 40dB. The filtered signal passes through a VIPA to separate spectral components that are detected by an sCMOS camera. (B) Pixel location in the spectrum are converted into frequency (see details in §2.2) to identify the Brillouin frequency shift $\omega_b$ after the peaks are fitted by a Lorentzian function, where $\omega_b = (\text{FSR} - \Delta f)/2$. FSR, full spectral range. $f(x_i)$ represents the frequency at pixel location $x_i$, as needed for Equation S8.

*Supplemental Information*

## 2.2 Data analysis

The Brillouin modulus $M$ was calculated following measurements of the Brillouin frequency shift $\omega_b$, refractive index $n$, and density $\rho$, according to:

$$M = \rho \left( \frac{\lambda \, \omega_b}{2 \, n \, \sin\frac{\theta}{2}} \right)^2 \tag{S7}$$

where $\lambda$ and $\theta$ are the wavelength and angle between the incident and scattered wave vectors, taken to be 561 nm and $\pi$, respectively. The Brillouin frequency shift was measured based on the frequency difference between the Rayleigh and Stokes peaks, as described below. Density was calculated according to Equations S1 or S5. Refractive index was measured using an Abbe refractometer (Bellingham and Stanley Ltd., London). For PEO hydrogels, a small sample (~0.05 ml) was taken and placed in the refractometer for measurement. For PA hydrogels, thin films were prepared from the same stock solution for each cross-linker concentration, allowed to swell in distilled water, and samples were cut out for refractometry at each time point.

Raw images of spectra were calibrated to convert pixel location to frequency. To do this, we identified the line in the sCMOS image connecting the Rayleigh, Stokes, and anti-Stokes peaks across two orders. A peak finding algorithm was performed to locate the pixel location of the six peaks $x_j$ along this line, where $j$ ranges from 1 to 6 (Figure S2B). A moving average was used to reduce noise. The position of each peak was then used to map pixel location to frequency $f$ based on the known free spectral range (FSR) of the VIPA according to $f = ax^2 + bx + c$, where $a$, $b$ and $c$ were determined by minimizing the least squared error of the following relationships:

$$\begin{bmatrix} f(x_2) \\ f(x_5) \\ f(x_4) - f(x_1) \\ f(x_6) - f(x_3) \\ \bigl(f(x_3) - f(x_2)\bigr) - \bigl(f(x_2) - f(x_1)\bigr) \\ \bigl(f(x_6) - f(x_5)\bigr) - \bigl(f(x_5) - f(x_4)\bigr) \\ \bigl(f(x_6) - f(x_4)\bigr) - \bigl(f(x_3) - f(x_1)\bigr) \end{bmatrix} = \begin{bmatrix} 0 \\ FSR \\ FSR \\ FSR \\ 0 \\ 0 \\ 0 \end{bmatrix} = \begin{bmatrix} x_2^2 & x_2 & 1 \\ x_5^2 & x_5 & 1 \\ x_4^2 - x_1^2 & x_4 - x_1 & 0 \\ x_6^2 - x_3^2 & x_6 - x_3 & 0 \\ x_3^2 - 2x_2^2 + x_1^2 & x_3 - 2x_2 + x_1 & 0 \\ x_6^2 - 2x_5^2 + x_4^2 & x_6 - 2x_5 + x_4 & 0 \\ x_6^2 - x_4^2 - x_3^2 + x_1^2 & x_6 - x_4 - x_3 + x_1 & 0 \end{bmatrix} \begin{bmatrix} a \\ b \\ c \end{bmatrix} \tag{S8}$$

where $x_1$, $x_2$ and $x_3$ are the pixel locations of the Stokes, Rayleigh and anti-Stokes peaks of the first order, and $x_4$, $x_5$ and $x_6$ are the pixel locations of the Stokes, Rayleigh and anti-Stokes peaks of the second order. The pixel locations match those identified in Figure S2B.

Lorentzian functions were then fitted to the Stokes peak of the second order and the anti-Stokes peak of the first order, and the frequency difference between these peaks, $\Delta f$, was calculated (Figure S2B). The Brillouin frequency shift $\omega_b$ was then calculated according to:

$$\omega_b = \frac{1}{2}(FSR - \Delta f) \tag{S9}$$

*Supplemental Information*

## 2.3 Comparison to the Theoretical Model

To determine the best fit relationship between $M$ and $\varepsilon$, Equation 2 was rearranged to yield

$$M = \frac{M_f M_s}{M_f + \varepsilon(M_s - M_f)} \tag{S10}$$

We then fitted Equation S10 to the data shown in Figures 1B and 2C of the main text, allowing $M_s$ and $M_f$ to be free parameters. We used a non-linear least squares Levenberg-Marquardt fitting algorithm in Matlab. The same fitting approach was applied to Figures S6B&E.

For the PEO data shown in Figure 1D, fits of $1/M$ versus $\varepsilon$ were applied using Equation 2 directly. We measured $M_f$ for each solvent in the absence of PEO by Brillouin microscopy, averaging over 3 locations with 100 measurements per location for each solvent. Equation 2 was then fitted to the remaining Brillouin measurements to estimate the value of $M_s$ that minimizes the root mean-squared error (RMSE) between Equation 2 and measured values of $M$. RMSE was calculated according to

$$RMSE = \sqrt{\frac{1}{N}\sum\left(\frac{1}{M} - \frac{1}{M_s} - \varepsilon\left(\frac{1}{M_f} - \frac{1}{M_s}\right)\right)^2} \tag{S11}$$

where the summation occurs over the total number $N$ of Brillouin measurements represented by the data points in Figure 1D. The minimum value of RMSE occurred at $M_s$ = 16.35 GPa (Figure S3). Predictions of $1/M$ from Equation 2 were then plotted versus $\varepsilon$ using the value of $M_s$ given above and the measured value of $M_f$ for each saline concentration to yield the tracings shown in Figure 1D of the main text.

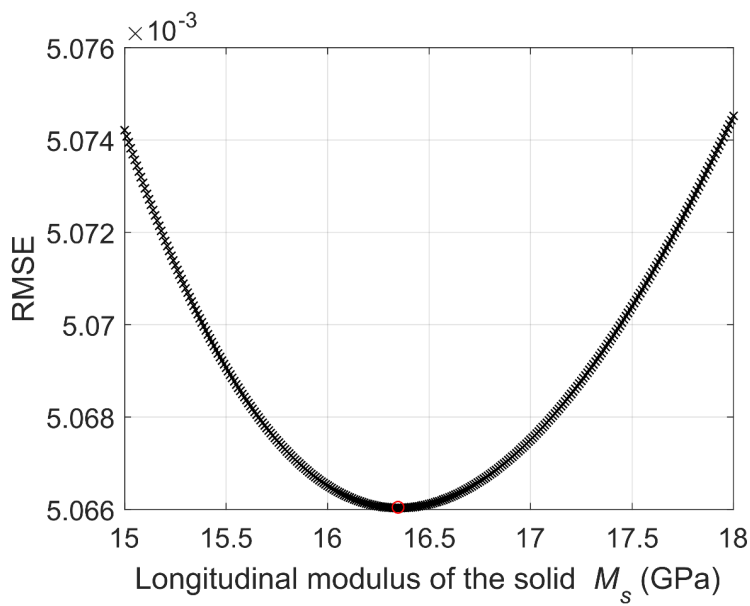

**Figure S3:** RMSE calculated according to Equation S11 as a function of the longitudinal modulus $M_s$ of the PEO polymer, representing the solid contribution of the hydrogel to the aggregate longitudinal modulus $M$. The minimum RMSE occurs at $M_s$ = 16.35 GPa. Data corresponds to Figure 1D of the main text.

## 3. Measurement of Young's modulus

### 3.1 PEO hydrogels

Measurements were performed using a controlled shear rate rheometer (AR 2000; TA Instruments) and Rheology Advantage software (TA Instruments). The rheometer used a cone plate configuration with a 40 mm diameter and 2° cone angle. Excess hydrogel was removed after lowering the cone. We first measured the viscoelastic storage modulus $G'$ at 1 Hz over a range of oscillatory strain (Figure S4A). These measurements confirmed that 1% strain lies within the linear elastic range. We then measured $G'$ and the viscoelastic loss modulus, $G''$, over a range of frequencies at 1% strain (Figure S4B). Assuming incompressibility, the Young's modulus was calculated as $E = 3G'$, where $G'$ was measured at 1 Hz and 1% strain. All measurements were done at 25°C.

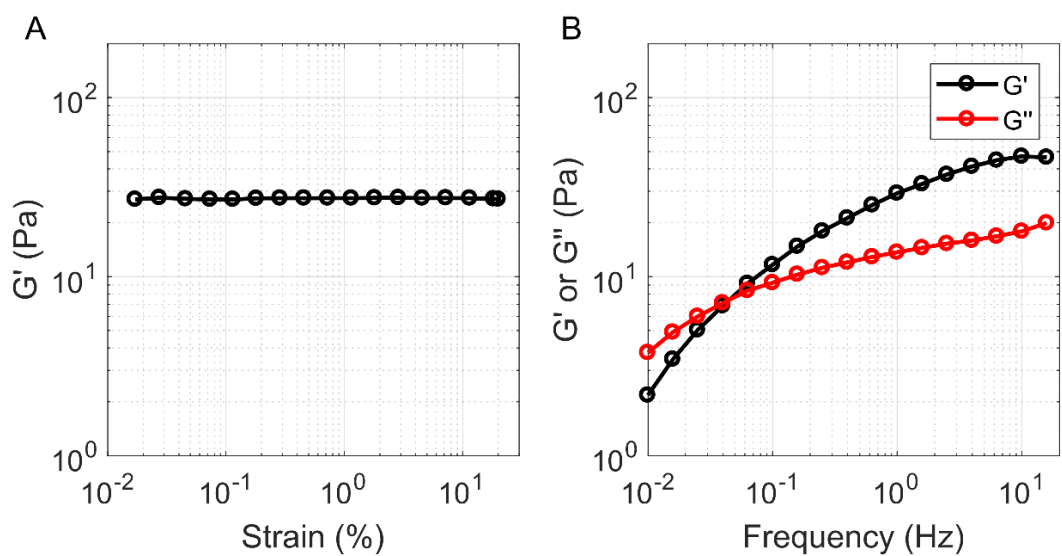

**Figure S4:** Representative rheometry data from PEO hydrogels showing the viscoelastic storage modulus ($G'$) and loss modulus ($G''$) as a function of oscillatory strain magnitude (A) and frequency (B). For this sample, the molecular weight was 8 MDa with $\varepsilon$ = 1.5%.

### 3.2 PA hydrogels

The instantaneous elastic modulus of PA hydrogels was measured in unconfined uniaxial compression using an Instron Model 5866 fitted with a 50 N load cell. The samples were preloaded to 0.01-0.05 N for one minute, then compressed to 0.7 mm at a crosshead speed of 0.5 mm/min for 2 or 3 cycles. Digital calipers with a precision of 0.01 mm were used to measure the thickness of the hydrogel and its diameter prior to compression, from which the area was calculated. Stress and strain were then calculated, and the compression modulus was obtained from a linear regression applied to the second or third cycle of the stress-strain curve (Figure S5). All measurements were done at room temperature.

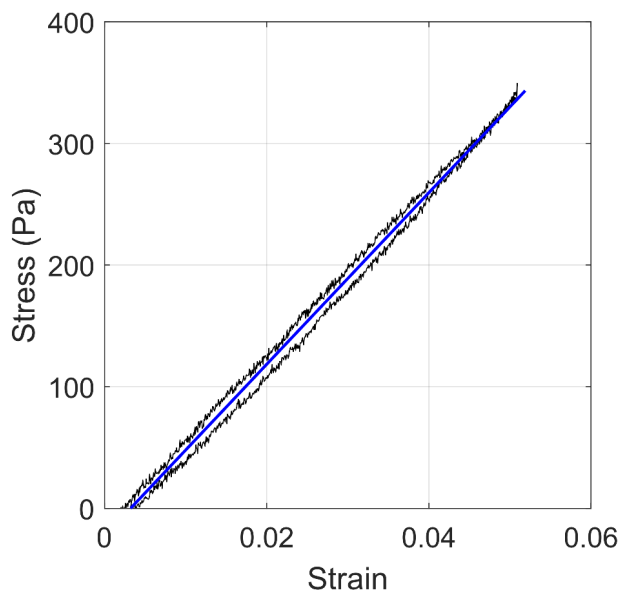

**Figure S5:** Representative stress-strain measurements (black tracing) obtained from a PA hydrogel in unconfined uniaxial compression. Linear regression (blue line) was applied to the complete cycle to calculate the Young's modulus as the slope of the regression line. The initial bis-acrylamide concentration of this sample was 0.06%. Data obtained after 12 hrs of swelling.

## 4. Code Availability

The custom Matlab code to analyze the data is available on request.

## 5. Data Availability

The data sets generated and analyzed during the current study are available on request.

## References for Methods

1. Tse, J.R. & Engler, A.J. Preparation of hydrogel substrates with tunable mechanical properties. *Current protocols in cell biology / editorial board, Juan S. Bonifacino ... [et al.]* **Chapter 10**, Unit 10 16 (2010).

2. Lepert, G., Gouveia, R.M., Connon, C.J. & Paterson, C. Assessing corneal biomechanics with Brillouin spectro-microscopy. *Faraday Discussions* **187**, 415-428 (2016)

3. Wilson, T & Sheppard C.J.R: *Theory and practice of scanning optical microscopy* London: Academic Press, 1984.

*Supplemental Information*

# Supplemental Notes

**Supplemental Note 1**

The data shown in Figures 1 and 2 of the main text were replicated in a separate set of experiments. PEO and PA hydrogels were prepared and analyzed as described in Supplemental Methods. PEO hydrogels included 1, 4 and 8 MDa with $\varepsilon$ = 0.92, 0.94, 0.96 and 0.98. Young's modulus for PEO hydrogels $E$ was measured by rheometry, as described in Supplemental Methods §3.1. PA hydrogels included 0.03, 0.06, 0.10, 0.15, 0.23 or 0.30% (w/v) bis-acrylamide and 10% (w/v) N-methylene-bis-acrylamide initial concentration, measured at 12, 30, 49 and 65 hrs after polymerization. For these studies, PA hydrogels were polymerized in disk-shaped molds of 20 mm diameter and 5 mm thickness. Young's modulus $E$ for PA hydrogels was measured by unconfined uniaxial compression, as described in Supplemental Methods §3.2, with a maximum displacement of 0.25 mm. The longitudinal modulus $M$ was measured by Brillouin microscopy, as described in Supplemental Methods §2. For the PEO studies, the VIPA spectrometer had a resolution of 0.4 GHz, a finesse of 70 and an FSR of 30 GHz. For the PA studies, the VIPA spectrometer had a resolution of 0.7 GHz, a finesse of 56 and an FSR of 39 GHz.

This second set of experiments exhibited the same behavior shown in Figure 1 and 2 of the main text. Briefly, $E$ for PEO hydrogels increased with increasing molecular weight for any given value of $\varepsilon$ due to increased entanglement between polymer chains (Figure S6A). There was no clear relationship between $M$ and $E$ (Figure S6C), but all values of $M$ collapsed onto a single curve predicted by Equation 2 when plotted versus $\varepsilon$ (Figure S6B). For PA hydrogels, $E$ decreased with increasing $\varepsilon$ (Figure S6D). No single relationship could describe $M$ in terms of $E$, but Equations 2 and 3 predicted the trajectory of how $M$ and $E$ changed during swelling (Figure S6F). Despite different swelling times and cross-linker concentrations, all values of $M$ collapsed onto a single relationship that depended on $\varepsilon$, as predicted by Equation 2 (Figure S6E).

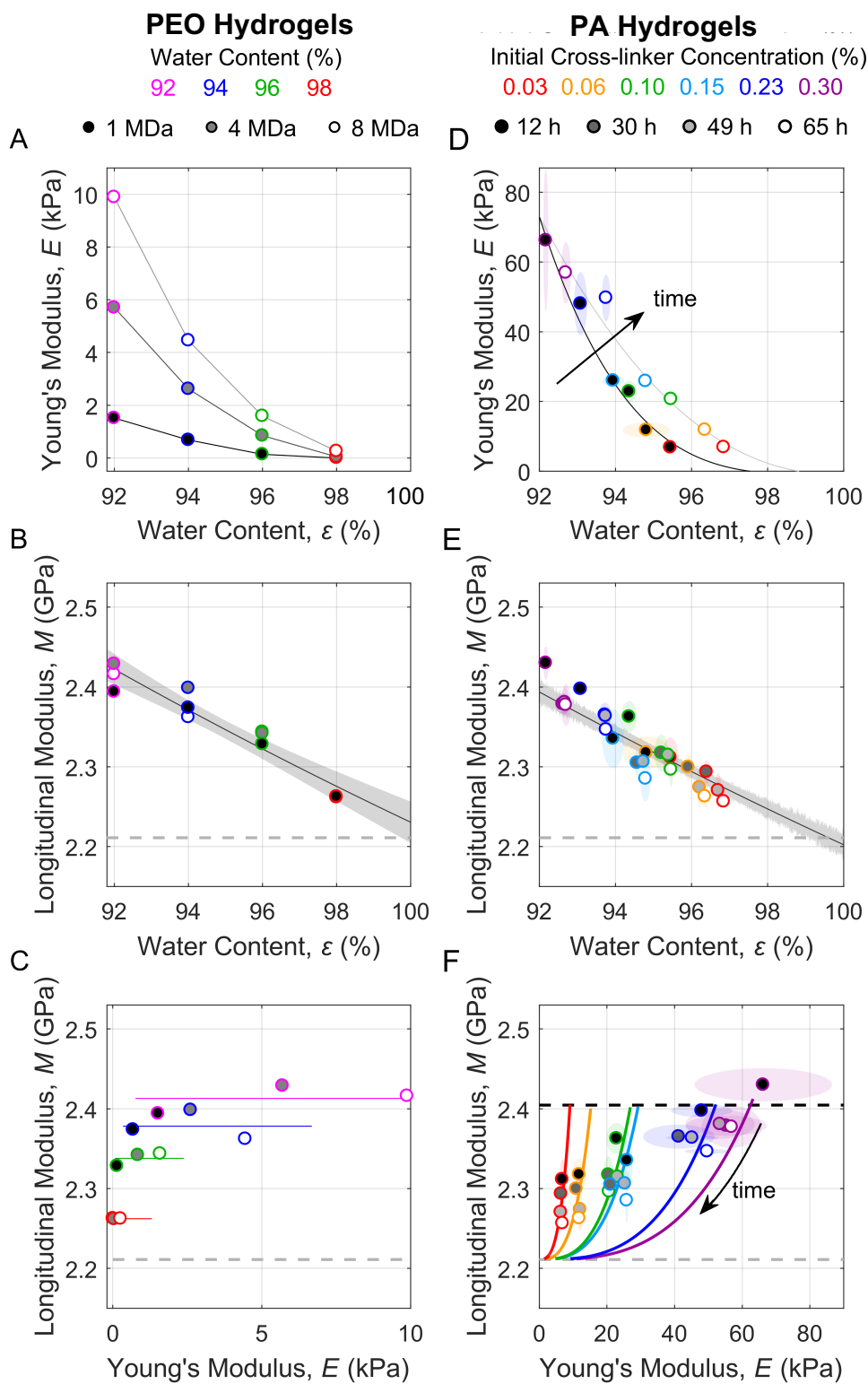

**Figure S6:** Replicate experiments for PEO (A-C) and PA (D-F) hydrogels give similar results as shown in Figures 1 and 2 of the main text. For PEO hydrogels, colors and shading represent water content and molecular weight, respectively, while for PA hydrogels colors and shading represent initial cross-linker concentrations and swelling times. These figures otherwise follow the format described for Figures 1 and 2 of the main text.

*Supplemental Information*

**Supplemental Note 2**

Previous studies have expressed the longitudinal modulus $M$ of hydrated materials measured by Brillouin microscopy in terms of the Young's modulus $E$ according to the empirical relationship:

$$\log(M) = a \log(E) + b \quad \quad \quad \text{(S12)}$$

where $a$ and $b$ are coefficients[1-5]. To examine whether this relationship exists in our data, we fitted Equation S12 to the PA data from Figure 2D. The analysis was applied separately to each time point during swelling and to all time points lumped together (Figure S7). For each case, there was a linear relationship between $\log(M)$ and $\log(E)$, and our estimated values of $a$ and $b$ were consistent with previously reported values (cf. $a$ = 0.023 and $b$ = 9.26[†] from Reference 4). We next examined whether the correlation between $M$ and $E$ arose from a direct relationship or rather could be explained by a mutual dependence on water content.

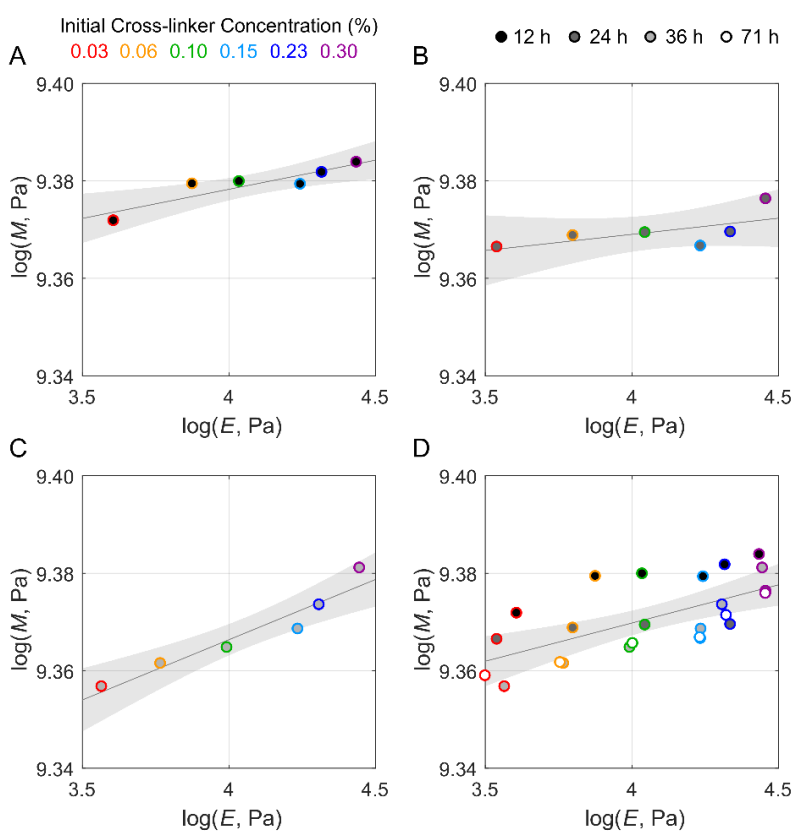

**Figure S7:** Longitudinal modulus $M$ measured by Brillouin microscopy versus Young's modulus $E$ plotted on logarithmic axes for PA hydrogels. Lines represent best fits of Equation S12 with $a$ = 0.012, $b$ = 9.33 for 12 h (A; $R^2$ = 0.82); $a$ = 0.007, $b$ = 9.34 for 24 h (B; $R^2$ = 0.41); $a$ = 0.025, $b$ = 9.27 for 36 h (C; $R^2$ = 0.91); and $a$ = 0.016, $b$ = 9.31 for all time points (D; $R^2$ = 0.43). Shaded regions show 95% confidence bounds. The color of marker outlines represents the initial bis-acrylamide concentration. Shading within the markers represent different swelling times. Data are taken from Figure 2D of the main text.

---

[†] The coefficients reported by Reference 4 were determined for $\log(M)$ versus $\log(G)$, where $G$ is the shear modulus. Because $E = 3G$, then $\log(E) = \log(G) + \log(3)$. The value of $b$ from Reference 4 was therefore corrected according to $b = b' - a' \log(3)$, where $a' = 0.023$ and $b' = 9.27$ (cf. Supplemental Note 7 from Reference 4).

*Supplemental Information*

With increasing water content $\varepsilon$, there was a decline in $E$ for both PEO (Figure 1A) and PA (Figure 2A) hydrogels. Similarly, with increasing $\varepsilon$, there was a decrease in $M$ (Figures 1B and 2C). Thus, as both $M$ and $E$ decrease with $\varepsilon$, a correlation between $M$ and $E$ may arise without a direct dependence of $M$ on $E$. This may be a general outcome, as many microstructural models (e.g., open cell foams[6], rubber elasticity[7], fiber-reinforced composites[8]) predict that stiffness decreases with increasing water content (decreasing solid fraction). Furthermore, Equation 2 predicts that $M$ decreases with increasing $\varepsilon$ whenever $M_s > M_f$. Thus, any analysis of the relationship between $M$ and $E$ must account for their mutual dependence on $\varepsilon$.

To control for $\varepsilon$, we asked whether Equation 2 provides any additional information on $E$ based on measurements of $M$ after accounting for the influence of water content. To do this, we subtract the value of $M$ predicted by Equation 2 from the value of $M$ measured by Brillouin microscopy, yielding $M_\varepsilon$, representing a hydration-corrected longitudinal modulus

$$M_\varepsilon = M - \frac{M_f M_s}{M_f + \varepsilon(M_s - M_f)} \tag{S13}$$

where $M_f$ and $M_s$ are the longitudinal modulus of the fluid and solid components, taken to be 2.21 and 16.35 GPa, respectively. For PA hydrogels at each time point during swelling or for all time points lumped together, there was no consistent correlation between $M_\varepsilon$ and $E$ (Figure S8). Thus, after correcting for water content, no obvious residual correlation remains between $M$ and $E$.

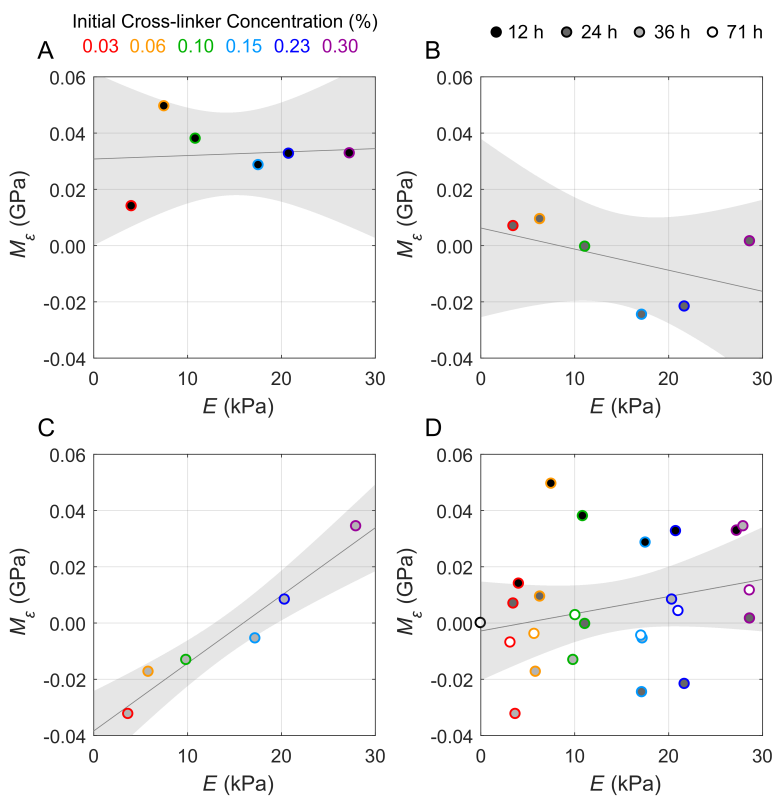

**Figure S8:** Hydration-corrected longitudinal modulus $M_\varepsilon$ versus Young's modulus $E$ for PA hydrogels. Lines represent best linear fits between $M_\varepsilon$ and $E$ after 12 h (A; $R^2$ = 0.01), 24 h (B; $R^2$ = 0.24) and 36 h of swelling (C; $R^2$ = 0.94) and all time points (D; $R^2$ = 0.06). $M_f$ and $M_s$ were taken to be 2.21 and 16.35 GPa, respectively. Shaded regions show 95% confidence bounds. The color of marker outlines represents the initial bis-acrylamide concentration. Shading within the markers represent different swelling times. Data are taken from Figure 2D of the main text.

*Supplemental Information*

The above analysis can be repeated for PEO hydrogels. However, as apparent from Figures 1C and S6C, variations in $E$ occur without detectable changes in $M$ when $\varepsilon$ is held constant. This demonstrates directly that the correlation between $M$ and $E$ largely vanishes after correcting for water content for PEO hydrogels.

This analysis demonstrates that for PEO and PA hydrogels any measurable change in $M$ is largely attributable to changes in $\varepsilon$, with little to no apparent influence of $E$ itself upon $M$. Any correlation observed between $M$ and $E$ for other hydrated materials, such as cells or tissues, may similarly arise due to a mutual dependence on $\varepsilon$, rather than a direct relationship between $M$ and $E$ per se. Depending on the microstructure, it is possible to obtain significant changes in $E$ without any detectable changes in $M$, as for the case of PEO hydrogels (Figure 1C). It is thus more appropriate to interpret measurements of Brillouin microscopy as a change in longitudinal modulus or local compressibility, which may occur independently of Young's modulus.

**References for Supplemental Note 2**